\documentclass[fleqn,twoside]{article}
\usepackage{espcrc2}

\usepackage{graphicx}
\usepackage[figuresright]{rotating}

\newcommand{\AmS}{{\protect\the\textfont2
  A\kern-.1667em\lower.5ex\hbox{M}\kern-.125emS}}

\newcommand{\vev}[1]{\langle #1 \rangle}

\title{Lattice QCD with dynamical domain wall quarks}

\author{Taku Izubuchi\address {Physics Department, Brookhaven
National Laboratory, Upton, NY 11973, USA}
\thanks{on leave from Institute of Theoretical Physics, 
Kanazawa University,   Ishikawa 920-1192, Japan.} 
\thanks{This manuscript has been authored under the support from 
Fellowship for Research Abroad of Japan Society for the Promotion of
Science.}
[RBC collaboration]
\thanks{%
The current members of the RBC collaboration are: 
Y.~Aoki, T.~Blum, N.~Christ, M.~Creutz, C.~Dawson, 
T.~Izubuchi, L.~Levkova, X.~Liao, G.~Liu, R.~Mawhinney, 
Y.~Nemoto, J.~Noaki, S.~Ohta, K.~Orginos, 
S. Prelovsek, S.~Sasaki and A.~Soni.
We thank  RIKEN, Brookhaven National Laboratory, and 
U.S. Department of Energy for providing the facilities 
essential for the completion of this work.}
}
\begin{document}

\begin{abstract}
We study lattice QCD with two flavors of dynamical domain wall 
quarks.  With renormalization group motivated actions, we 
find chiral symmetry can be preserved to a high degree at lattice cut off of 
\(a^{-1}\) \(\sim\) 2 GeV even for  fifth dimension size as small as
\(L_s=12\).  In addition two new steps are introduced to improve 
the performance of the hybrid Monte Carlo simulation.
\vspace*{-0.4cm}
\end{abstract}

\maketitle

\section{Introduction}

  An obvious source of systematic error in numerical lattice QCD at 
present is the quenched approximation.  Use of domain wall fermion 
(DWF) \cite{DWFbasics} dynamical quarks to lift this 
approximation is attractive because 
DWF preserve the flavor and chiral symmetries both on and off shell 
on the lattice before the continuum limit is taken.  Evidently successful 
implementation would bring many advantages, such as detailed study of 
flavor-singlet meson spectrum and its relation to the \(U(1)_A\) 
anomaly.  However some recent exploratory calculations \cite{CUtherm}
on  coarse lattices with cut off of \(a^{-1}\) \(<\) 1 GeV
show the chiral symmetry is rather poorly realized: impractically large 
 \(L_s\) \(\sim\) 100 are necessary to  bring the residual symmetry
violation under control.  In contrast 
in this work we go to a higher cutoff of \(a^{-1}\) \(\sim\) 2 GeV with 
renormalization-group (RG) motivated gauge actions and find that the 
residual violation can be made sufficiently small even at 
\(L_s =\) 12.  

  To cope with considerably larger demand for computing 
resources brought by this new approach, we introduce two improvements in 
the hybrid Monte Carlo (HMC) algorithm which accelerate 
the simulation reported below on the QCDSP by about a factor of three.

\section{New algorithms}
  The lattice QCD partition function with two flavors ($N_F=$2) of DWF quarks is
\begin{equation}
Z = \int {\cal D} U
{\det(D^\dagger D)|_{m=m_f} \over \det(D^\dagger D)|_{m=1} }
e^{-S_g},
\label{eq:DynDWF_PF}
\end{equation}
where $D$ is the 5d Dirac operator 
with 4d bare fermion mass $m$ and $S_g$ the gauge action.
The determinant in the denominator in Eq. \ref{eq:DynDWF_PF} comes from 
the Bosonic Pauli-Villars regulators, $\phi_{PV}$, whose mass is set to
$1$  to cancel the dominating contribution from the fermions on 
the bulk 5d lattice.
This allows us to study 4d QCD with 2 flavors of light quarks
localized at the boundary and coupled to the 4d gauge
field. The original fermion action employed in \cite{PavlosSchw,CUtherm} was
\begin{eqnarray}
S_F = \sum_{x,s}  \{-\phi_F^\dagger (D^\dagger D)|_{m=m_f}^{-1}\phi_F\\ \nonumber
\ \ \ \ \ \  \ \ \ \ \ \ \ \ \ \ \ -\phi_{PV}^\dagger(D^\dagger D)|_{m=1}\phi_{PV}\},
\end{eqnarray}
where $\phi_F$  is the pseudo-fermion field.
By integrating out $\phi_F$ and $\phi_{PV}$, we obtain 
the partition function Eq. \ref{eq:DynDWF_PF}.
We chose  hybrid Monte Carlo phi algorithm (HMC-$\phi$) \cite{HMC} 
to generate the Markov chain of gauge configurations.

Noticing equivalence between 
\(\frac{\det (A^\dagger A)}{\det (B^\dagger B)}\) and 
\(\{\det [B (A^\dagger A)^{-1} B^\dagger]\}^{-1}\),
we could also obtain the partition function Eq. \ref{eq:DynDWF_PF}
from a new action
\footnote{
This 
part of the work is done mainly by C.~Dawson.
}
\begin{equation}
S' = -\phi_F^\dagger[ D|_{m=1} (D^\dagger D)|_{m=m_f}^{-1} D|^\dagger_{m=1}] \phi_F
\label{eq:newaction}
\end{equation}
where we need only one bosonic field $\phi_F$. 
Even/odd preconditioning is implemented in either case.

The benefits of the new action and the associated force term are
first the unwanted contribution from 5d bulk fermion is 
now canceled explicitly while it was canceled only stochastically in the
original force term. This leads to better acceptance in the 
Metropolis accept/reject step, which  increased by about 20\% in a
realistic simulation.
Secondly, the source vector, $\phi_F$, of the Dirac equation solved in 
the force term calculation is modified to $D^\dagger|_{m=1}\phi_F$ as seen
in Eq. \ref{eq:newaction}, and it turns out that
the number of conjugate gradient(CG) iterations is 
20 - 30 \% smaller than that of the original action.


The second improvement \cite{ChronoOrder} is to
forecast the solution of the CG at each step of a HMD trajectory 
from the solutions of previous steps.
In each of trajectories, the system obeys the smooth 
classical molecular dynamics (MD), so future solutions can be estimated
from past ones.
We choose the Minimal Residual Extrapolation  method of
\cite{ChronoOrder} with a repeated Gram-Schmidt orthogonalization
to ensure orthogonality. The  solution is forecasted by minimizing
the norm of the residual vector within the subspace spanned by previous solutions. 
Then the forecasted solution is passed into the ordinary CG 
as an initial input vector.

\begin{figure}
\includegraphics[scale=0.3]{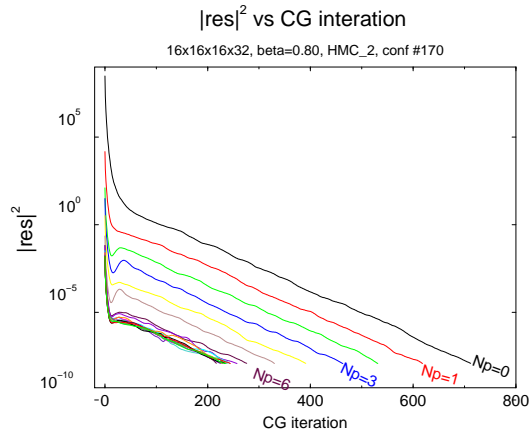}
\vspace*{-0.80cm}
\caption{The squared norm of the residual vector 
as  a function of CG steps. 
The initial solution vector is  precognized
based on previous $N_p$ solutions.}
\vspace*{-0.80cm}
\label{fig:precog_cg_2}
\end{figure}

The manner of convergence of the CG based on different numbers of 
previous solutions in the forecasting  is shown in
Fig. \ref{fig:precog_cg_2}. 
The squared norm of the residual vector 
of a solution (vertical axis) is  already small, ${\cal O}(1)$, at zero  CG count
(horizontal axis)  when  the number of previous solution used in the forecast ($N_p$) 
is larger than one. 
The 
CG iterations required for 
convergence decreases monotonically with increasing $N_p$, up to 7.  

The reversibility required to prove the detailed balance in the Markov chain
is checked by flipping the sign of the conjugate momentum field at 
the end of a trajectory and evolving the configuration backward by the same
number of MD steps.
The resulting configuration obtained at the end of the reversed process 
differs from the initial configuration  
by less than $2 \times 10^{-5}$ in  elements of SU(3)
link variables for a CG convergence criteria, $|res|/|src| < 10^{-8}$. 
By this trick with $N_p=7$ the number of total CG count is  55 \% of 
the original code (using $N_p=1$) and  about 35 \% of a simulation
without the forecasting.

\section{Physical scale of the dynamical lattice}
  Observing the large chiral symmetry violation for coarser lattice
spacing, $a^{-1} < 1$ GeV \cite{CUtherm}, we decided to perform a large scale
simulation with target $a^{-1}\sim 2$ GeV.
To quickly determine  parameters for this target lattice spacing,
we examined the parameter space on an $8^3\times 4$ lattice
and compared the meson spectrum to that obtained with the dynamical 
staggered fermions on the same lattice size assuming that scaling violations
from the two fermion actions are mild.
We also explored the dependence of the residual chiral symmetry breaking
with various RG improved gauge actions.
We found at fixed $a^{-1}\sim 2$ GeV that  chiral symmetry 
was better realized 
for larger negative values of the coefficient of the $(1\times2)$
rectangular plaquette although the dependence of 
the residual mass, $m_{res}$, was much more mild than for quenched
simulations \cite{DBW2}. 

  For our most intensive simulation, we chose  DBW2
gauge action at $\beta=0.80$ with $N_F=2$ on a $16^3\times32$ lattice. 
The sea quarks have mass  $a m_f^{(dyn)}=0.02$,
domain wall height $M_5=1.80$, and $L_s=12$.
An 
trajectory  consists of 50 leap-frog steps
and advances  the MD time, $\tau$, by 0.5.

  The average number of CG iterations to fulfill
the convergence criteria of $|res|/|src| < 10^{-8}$ is 550 for $N_p=1$ and
314 for $N_p=7$.
The acceptance ratio of the accept/reject step is about 77(3) \%.
On a 4096 node partition ($\sim$ 205 GFLOPS peak) of the QCDSP, it takes about
1 hour per trajectory. 

  To estimate the lattice spacing within the statistics
accumulated so far,  we measure the heavy 
quark potential and the meson spectrum
from  the last 50 lattices which are separated by five
trajectories after dropping the first 500 trajectories.
We monitor the chiral condensate $\vev{\bar q q}$ as a function of 
the trajectory number and pick lattices for measurement 
after  $\vev{\bar q q}$ is considered to have reached its thermalized value.
However, we note that it is not clear whether the evolution 
is thermalized at this point. 

\begin{figure}
\includegraphics[scale=0.3]{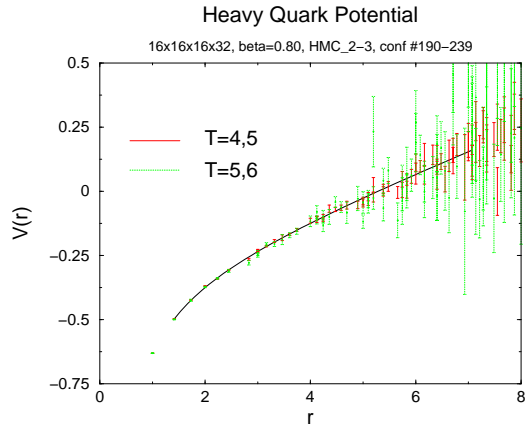}
\vspace*{-0.80cm}
\caption{The heavy quark potential obtained from $T=4-5$ (solid error
 bars) and $T=5-6$ (dotted error bars) as a function of distance between
quark and anti-quark with its fitted curve.}
\label{fig:pot}
\vspace*{-0.60cm}
\end{figure}

 Measuring  the correlation between path ordered products of gauge-fixed
links in the time direction of length $T$ at a space point 
$x$, $L_T(x)$, we obtain the heavy quark potential 
$ e^{-V(r)T} = \vev{L^\dagger_T(x)L_T(x+r)}$
\footnote{
This calculation is
done using the MILC code, version 6 for which we thank  MILC collaboration.}.
As shown in Fig. \ref{fig:pot}, 
the fit to the data for $4 \le T \le 6$ and $\sqrt{2} \le r \le 7$ to  
the function $V(r)=C -\alpha/r + \sigma r$  gives 
$a\sqrt\sigma=0.28(2)$, or $a^{-1} \approx$ 1.7(2) GeV, 
and the Sommer scale, $r_0/a = 4.1(3)$.

 Meson propagators are calculated using  valence quarks with 
$L_s=12, M_5=1.8$ and $a m_f^{(val)}=0.015, 0.020, 0.025, 0.030$.
Extrapolating the vector meson mass linearly,  
we find $aM_V(m_f^{(val)}\to0)=$0.34(9)
which is consistent with the results of the heavy quark
potential within (large) error.  
The squared mass of the pseudo-scalar meson is well fit 
to the PCAC relation. 
Taking the Kaon mass from experiment and the lattice spacing 
from the heavy quark
potential yields a sea quark mass approximately half the strange quark mass.

Finally, the residual chiral symmetry breaking as determined from the
pseudo-scalar mid-point correlator \cite{DWF_spect}
turns out to be $a m_{res}(m_f\to0)=1.47(7)\times 10^{-3}$, 
which  corresponds to a few MeV.


\section{Conclusion and Discussion}
In summary, we have begun investigating lattice QCD with 
two  flavors of dynamical domain wall
quarks.  Assuming that the lattices have reached 
thermal equilibrium,
we could interpret the configurations as QCD at roughly 0.1 fm lattice
spacing, sea quarks with mass about $m_s/2$, and
residual chiral symmetry breaking of only a few MeV. 
This level of symmetry breaking is now commensurate to that observed 
in quenched studies.

 Two new algorithms were introduced to enhance the performance 
of the simulation.
The speed up for the parameters discussed in this pilot study
is about a factor of three. 
With this acceleration, further
developments of theoretical and numerical techniques, and more
computing horse power \cite{QCDOC}, we may reach our goal
to study physics in the continuum limit of QCD.

\end{document}